\documentclass{article}

\usepackage{graphics}
\usepackage{float}
\usepackage{cite}
\usepackage{amsmath,amssymb,amsfonts}
\usepackage{graphicx}
\usepackage{textcomp}
\usepackage{algorithm}
\usepackage{algpseudocode}
\usepackage{booktabs}
\usepackage{siunitx}
\usepackage{caption}

\usepackage[final]{neurips_2025}

\usepackage[utf8]{inputenc} 
\usepackage[T1]{fontenc}    
\usepackage{hyperref}       
\usepackage{url}            
\usepackage{booktabs}       
\usepackage{amsfonts}       
\usepackage{nicefrac}       
\usepackage{microtype}      
\usepackage{xcolor}         

\title{Trading on Uncertainty: FutureQuant Transformer's Distribution-Based Strategy for Futures Markets}

\author{
    Wenhao Guo$^{*}$\\
    Department of Computer \\
    Science \& Engineering\\
    Ohio State University\\
    \texttt{guo.2484@osu.edu} \\
    \And
    Yuda Wang$^{*}$\\
    School of Computing\\
    and Data Science\\
    Hong Kong University\\
    \texttt{yuda\_wang@connect.hku.hk}\\
    \And
    Zeqiao Huang$^{*}$\\
    Tandon School of Engineering\\
    New York University\\
    \texttt{zh3194@nyu.edu}\\
    \And
    Changjiang Zhang$^{\dagger}$\\
    College of Science, \\
    Mathematics and Technology\\
    Wenzhou Kean University\\
    \texttt{chazhang@kean.edu} \\
    \And
    Shumin Ma$^{\dagger}$\\
    Faculty of Science and Technology\\
    Beijing Normal-Hong Kong Baptist University\\
    \texttt{shuminma@uic.edu.cn} \\
}

\begin{document}

\maketitle

\begin{abstract}
  In the complex landscape of traditional futures trading, where vast data and variables like real-time Limit Order Books (LOB) complicate price predictions, we introduce the FutureQuant Transformer model, leveraging attention mechanisms to navigate these challenges. Unlike conventional models focused on point predictions, the FutureQuant model excels in forecasting the range and volatility of future prices, thus offering richer insights for trading strategies. Its ability to parse and learn from intricate market patterns allows for enhanced decision-making, significantly improving risk management and achieving a notable average gain of 0.1193\% per 30-minute trade over state-of-the-art models with a simple algorithm using factors such as RSI, ATR, and Bollinger Bands. This innovation marks a substantial leap forward in predictive analytics within the volatile domain of futures trading.
  
  \textbf{Keywords:} Futures market, Time series forecasting, Quantile regression, Transformer, Market Risk, Trading Signal, Cumulative Return
\end{abstract}

\section{Introduction}

\subsection{Background}
The futures market plays an indispensable role in the global financial ecosystem, offering mechanisms for price volatility hedging and price discovery for commodities and financial instruments. The challenge of accurate prediction within this market stems from its inherent complexity, influenced by diverse factors such as economic indicators, geopolitical events, and market sentiment. Traditional prediction models, including ARMA, ARIMA, and GARCH, often fall short of capturing these dynamics due to their linear and static nature.

The emergence of machine learning, particularly deep learning techniques, marks a significant leap forward in financial forecasting. These models are distinguished by their ability to discern nonlinear relationships and adapt to evolving data, thereby improving prediction accuracy. A notable trend is the shift towards models forecasting distributional outcomes, such as future price quantiles, through quantile regression. This method offers a panoramic view of potential risk and variability, crucial for informed decision-making and risk management.

The incorporation of Long Short-Term Memory (LSTM) networks and attention mechanisms into financial models has refined their ability to process sequential data. This enhancement is pivotal for capturing long-term dependencies and identifying critical features. The introduction of the FutureQuant Transformer model, leveraging the self-attention capabilities inherent in transformer architectures, presents a promising avenue for unlocking deeper market insights.

This research aims to evaluate the effectiveness of cutting-edge machine learning models in futures market price prediction, with a special focus on quantile-based forecasts. These advanced models hold the potential to provide strategic advantages in navigating financial risks.
\subsection{Objectives and Contributions}
The impetus for this study stems from the limitations of the traditional futures market prediction methods, which mainly concentrate on price or trend forecasting. Although useful, their applicability is diminished by the market's complexity and fluidity. A recent and significant shift in financial market analysis is the move from single-point predictions to distribution predictions, acknowledging the superiority of understanding the full spectrum of potential market movements.

This shift towards distribution prediction represents a more holistic approach, projecting the entire price distribution landscape and thus furnishing a richer understanding of market dynamics and potential risks. Utilizing advanced machine learning techniques for distribution prediction heralds new possibilities for risk assessment and strategic decision-making, enabling a preparedness for various market scenarios beyond a singular forecast.

The contributions of this study to the field of futures market prediction are threefold:

\begin{enumerate}
  \item \textbf{Adaptation of Attention Techniques to Finance}: We innovatively adapt attention mechanisms, originally from NLP, to the financial context of futures trading. This addresses the challenges posed by the high-dimensional data of the limit order book, showcasing the versatility of NLP-inspired models in financial analysis.

  \item \textbf{Innovative Model Output for Risk Management}: By shifting the model's focus to predicting future price quantiles and distributions instead of mere point estimates, we enable a more nuanced approach to risk management. This adaptation offers traders a comprehensive dataset for more informed decision-making.

  \item \textbf{Strategic Trading Strategy Development}: The experimental validation of our model has led to the development of a trading strategy that integrates financial indicators such as RSI, Bollinger Bands, and ATR. This strategy, aligned with our model’s outputs, demonstrates enhanced short-term returns with mitigated risk, outperforming traditional models.
\end{enumerate}

These contributions collectively advance the frontier of financial market analysis, introducing a transformative approach that leverages the sophisticated capabilities of NLP-inspired models for a deeper exploration of the futures market data. The FutureQuant Transformer model, by providing distribution-based forecasts, equips traders with superior risk assessment tools and strategic insights, establishing a new benchmark in financial market prediction and underscoring the significant impact of advanced machine learning techniques in deciphering the complexities of futures trading.

\section{Related Work}

\subsection{Futures Market Prediction Methods}\label{AA}
The evolution of futures market prediction methodologies has transitioned from reliance on traditional statistical models, such as ARIMA, to integration of advanced computational techniques. These statistical models laid the foundational understanding of market trends but struggled to fully encapsulate the dynamic and complex nature of futures markets. The advent of deep neural networks, particularly for analyzing limit order books (LOB), marks a significant advancement. Convolutional Neural Networks (CNNs) have proven effective in capturing spatial structures within LOB data, offering a nuanced view of market dynamics [1]. A Further innovation is seen in the fusion of CNNs with Recurrent Neural Networks (RNNs), especially Long Short-Term Memory (LSTM) networks, enhancing the capacity to model long-range temporal dependencies and thereby elevating prediction accuracy [2].

The introduction of Transformer models to the analysis of LOB data signifies another leap forward. Originally devised for natural language processing tasks, Transformers excel in handling sequential data through self-attention mechanisms, enabling the modeling of complex patterns in LOB data that eluded traditional models[3]. This trajectory towards employing state-of-the-art AI methodologies in financial time series analysis marks a pivotal shift, potentially leading to unprecedented market prediction accuracy and efficiency

\subsection{The Importance of Distribution Prediction}
The burgeoning interest in utilizing high-frequency intra-day data to predict financial asset returns and volatilities reflects a broader ambition to understand asset risk distributions. The futures market, characterized by higher volatility compared to traditional assets, necessitates a nuanced approach to risk assessment. This research adopts quantile regression models to forecast future price distributions, leveraging realized measures as explanatory variables. This methodology underscores the value of non-parametric models in capturing intra-day price behaviors, facilitating refined trading strategies and effective risk management[4].

\subsection{ LSTM and Transformer}
\subsubsection{LSTM in Futures Market Prediction}
Long Short-Term Memory (LSTM) networks, Specialized in managing long-range dependencies within time series data, LSTM networks have found extensive application in financial market predictions, including futures markets. Their ability to leverage past information across long sequences sets them apart from traditional RNNs, as highlighted in the seminal work by Fischer and Krauss (2018) on stock market predictions, indicating broader applicability to financial market analysis[5].

\subsubsection{Transformer}
Revolutionizing NLP through the "Attention is All You Need" paper by Vaswani et al. (2017), the Transformer model's self-attention mechanism has been adapted for a variety of domains, including finance. Its efficacy in time series prediction and risk management underscores the model's versatility, marking a significant foray into futures market analysis[6].

\subsection{Estimating Predicting Intervals}

Consider two random variables $M \in \mathcal{M}$ and $N \in \mathcal{N}$, representing input observations and their corresponding outcomes, respectively. We denote the joint distribution of $M$ and $N$ as $\rho(M, N)$, and express the conditional distribution of $N$ given $M$ as $\rho(N \mid M)$.

A Prediction Interval (PI), formulated from a collection of training data $\{(m_j, n_j) \mid j = 1, 2, \ldots, m\}$, where each pair $(m_j, n_j)$ exemplifies an instance of $\rho(M, N)$, is characterized by $\hat{D}_{\rho,m}(m) = [G(m), H(m)]$. In this context, $G$ and $H$ map the input domain $\mathcal{M}$ to the outcome domain $\mathcal{N}$. The magnitude or span of a PI, computed as $H(M^*_{m+1}) - G(M_{m+1})$, is influenced by the confidence parameter $\beta$, indicating the probability $1 - \beta$ that a new sample will be contained within the PI. Broader intervals signify higher uncertainty in forecasts.

The efficacy of a PI is judged by its capacity to encompass the true value  being forecasted. A prediction interval is considered accurate if the coverage probability for a new data point $(M_{m+1}, N_{m+1}) \sim \rho$ is equal to or larger than the chosen confidence threshold. Further details on the assurances provided by PI coverage are discussed later.
    
\subsection{Coverage in Marginal and Conditional}
 PI coverage manifests in two principal forms: marginal coverage assurance, computed as an average across several test samples, and conditional coverage assurance, relevant to a specific fixed value $M_{m+1} = m$. The marginal coverage assurance, which remains neutral to the specific underlying distribution, posits that the likelihood of the PI including the actual test outcome $N_{m+1}$ should be no less than $1 - \beta$, calculated over an average from any base distribution $\rho$. This is mathematically articulated as:
 $$
P(N_{m+1} \in \hat{D}_{\rho,m}(M_{m+1})) \geq 1 - \beta \quad (1)
$$

\textbf{Conditional coverage}, in contrast, sets a more rigorous benchmark. A PI meets conditional coverage at the $1 - \beta$ level if, for any specified point $m$, the probability that $\hat{D}_{\rho,m}$ includes the point $M_{m+1} = m$ is at least $1 - \beta$. This is expressed as:
$$
P(N_{m+1} \in \hat{D}_{\rho,m}(M_{m+1}) \mid M_{m+1} = m) \geq 1 - \beta \quad (2)
$$

To show the distinction between conditional coverage and  marginal in the context of futures trading, envision a scenario where each data point $j$ symbolizes a unique market scenario, with $M_j$ representing key market indicators (like price movements, trading volumes, etc.), while $N_j$ denotes a particular market result (such as a change in price). In forecasting the future market scenario $M_{m+1}$, a trader or analyst seeks to predict the result $N_{m+1}$ within a certain range, stating: "Based on the current market indicators, the price is anticipated to vary by X to Y amount."


\subsection{Quantile Loss}
Quantile Regression (QR) primarily focuses on predicting a specific conditional quantile function (CQF) of the variable $Y$ which is given $X$, at a predetermined quantile level $\beta$. This CQF is formally denoted as:
$$
q_\beta(x) = \text{min}\{y \in \mathbb{R} \,|\, P_Y(Y \leq y \mid X = x) \leq \beta\} \quad (3)
$$

Here, $P_Y(y)$ represents the conditional cumulative distribution function of $Y$, which is essential for deriving the probability density function from empirical CQFs. These empirical CQFs are computed within the miscoverage range of $0 < \beta < 1$.

In QR, Prediction Intervals (PIs) are derived from two empirical CQFs obtained from the training data. The confidence measure, indicated by $1 - \beta$, denotes the difference between two selected quantile estimates. The estimated PI in QR is thus defined as:
$$
\hat{C}_\beta(x) = [\hat{q}_{\beta_{\text{lo}}}(x), \hat{q}_{\beta_{\text{hi}}}(x)] \quad (4)
$$

As $\hat{q}_{\beta_{\text{lo}}}(x)$ and $\hat{q}_{\beta_{\text{hi}}}(x)$ refer to the lower and upper empirical CQFs, calculated for quantiles $\beta_{\text{lo}} = \beta / 2$ and $\beta_{\text{hi}} = 1 - \beta / 2$, respectively. Unlike the fixed width of the PI in Equation (4), the breadth of the PI varies for each data point $x$, adapting to the heteroscedastic nature of the data.

But replacing the  interval $C_\beta(x)$ with the empirical  $\hat{C}_\beta(x)$ estimation from Equation (4) does not ensure that the actual PI coverage will consistently match the targeted confidence level $(1 - \beta)$.

The calculation of $\hat{q}_{\beta_{\text{lo}}}(x)$ and $\hat{q}_{\beta_{\text{hi}}}(x)$ is an optimization problem, focused on minimizing the pinball loss function. The pinball loss for a specific data combination  $(x_i, y_i)$ is defined as:
$$
L_{\beta, i} = \left\{
    \begin{array}{ll}
        (1 - \beta) \cdot (\hat{q}_\beta(x_i) - y_i) & \text{when } \hat{q}_\beta(x_i) \geq y_i \\
        \beta \cdot (y_i - \hat{q}_\beta(x_i)) & \text{when } \hat{q}_\beta(x_i) < y_i
    \end{array}
\right. \quad (5)
$$

Here, $y_i$ is the observed outcome for the sequence i instance, and $\hat{q}_\beta(x_i)$ is the estimated $\beta$-quantile. The quantile loss assesses the accuracy of the quantile estimate against the actual data, with a lower value indicating a more precise estimate. This loss metric is also crucial in the training of deep learning models.

\section{Research Methodology}
\subsection{Data Overview}

The study's primary objective is to develop a model for predicting futures prices and to establish an associated trading strategy. Achieving this goal necessitates the acquisition of real-time market data. To facilitate this, the researchers have designed a data retrieval application programming interface (API) using C++. This API enables the extraction of data concerning Chinese futures products from the Futures Exchange Center.

In adherence to regulations aimed at maintaining market stability, which restrict direct individual access to the futures trading system, the researchers established an official financial account with a registered securities firm. This arrangement provided the necessary transactional capabilities and access to information.

Utilizing the C++ API, the researchers requested specific data points, concentrating on the Ag2312 silver futures contract, which has a trade split date set for December 2023. The selected data attributes for collection are “UpdateTime,” “UpdateMillisec,” “LastPrice,” “Volume,” “BidPrice1,” “BidVolume1,” “AskPrice1,” and “AskVolume1.”

“UpdateTime” and “UpdateMillisec” serve as timestamps for the data capture, recording every half-second in reaction to any data changes. “LastPrice” indicates the price at which the latest trade of the futures contract occurred. “Volume” measures the transaction count for a specific futures contract within a set timeframe, acting as a vital indicator of market activity and liquidity.

“BidPrice1” and “BidVolume1” detail the maximum price buyers are prepared to pay and the corresponding trade volume at this bid price, respectively. In contrast, “AskPrice1” and “AskVolume1” denote the minimum price sellers are willing to accept and the trade volume at this ask price, respectively. These metrics offer crucial insights into the market depth and prevailing trends. A narrow bid-ask spread signifies a more liquid market, while a broader spread points to decreased liquidity and a higher potential for price volatility. Understanding these parameters is essential for traders and investors to analyze market conditions and make well-informed trading decisions.
\subsection{Model Comparison}
\subsubsection{Quantile LSTM}
\paragraph{Model Training with Quantile Loss}

The Quantile LSTM model is refined using a loss function aimed at precise predictions across designated quantiles of the future distribution of market prices. This loss function, known as the quantile loss and represented by $L_{q, \alpha}$, is essential for training the model to understand the conditional distribution of the target variable over a set of quantiles $\{\alpha\}$.

For each predicted output $\hat{q}_{\alpha}(x_i)$ pertaining to the $\alpha$-quantile, the quantile loss is computed as indicated in the specified equation. It asymmetrically penalizes prediction errors, which facilitates optimization across different segments of the predicted distribution. This feature is crucial for capturing the entire scope of possible future market behaviors.
\paragraph{Window Size Configuration} 
The LSTM model employs a specific window size of 30 time steps, denoted as $\text{window\_size\_in}$, to incorporate the relevant historical data needed for prediction. This window size is chosen to ensure a balance between including enough historical context and avoiding excess computational demands.

The model aims to predict a single future time step, indicated by $\text{window\_size\_out}$, concentrating on the $\text{LastPrice}$ feature. This setting is optimized to accurately project imminent market trends, which are of significant importance for developing trading strategies and managing risk in the financial domain.
\subsubsection{FutureQuant Transformer}
\paragraph{Basic Structure Of FutureQuant}

The tensor that our model processes is of the form [batch size, sequence length, features], where features are the number of components you are attempting to predict and the sequence length is the number of time steps in your sample. Features are 1 if our goal is to forecast a single thing, such as the closing price. To avoid confusion with batch size during model fitting, I will refer to batch size as "samples" while discussing tensor forms.
$$
X \in \mathbb{R}^{N \times T \times F}  \quad(6)
$$

For a single transformer encoder block, see Fig. 1. below. Because a “LayerNormalization” layer performs better than a “BatchNormalization” layer, the embedding is eliminated for user's needs. Positional embedding is one approach, but to be honest, in this paper, authors have tried three different time2vec implementations, and although the code functions, the outcomes are worse. Finally, the authors choose the best one to use.
$$ X_{\text{norm}} = \frac{X - \mu}{\sigma} \odot \gamma + \beta \quad(7)$$
The Multi-Head Attention is already coded for in Keras. The feedforward part can be almost anything.
$$\text{MultiHeadAttention}(Q, K, V) = \text{softmax}\left(\frac{QK^T}{\sqrt{d_k}}\right)V \quad(8)$$
Where Q,K,V in equation (8) are the query, key, and value matrices, respectively, and $d_k$ is the dimension of the key vectors.
Conv1D layers (because results are good) could get away with almost anything, dense layers, another LSTM, whatever you want. 
$$Conv1D(LayerNorm(X))=ReLU(Conv1D(X)) \quad(9)$$
Where Conv1D in equation (9) is a one-dimensional convolutional layer and ReLU is the Rectified Linear Unit activation function.
\begin{figure}[htp]
    \centering
    \includegraphics[width=5cm, height=10cm, keepaspectratio]{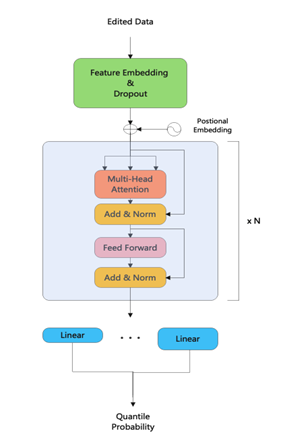}
    \caption{basic structure of the Quartile Transformer}
    \label{fig:galaxy}
\end{figure}
Multiple transformer encoder blocks can be stacked, which can also add the last Multi-Layer Perceptron/DNN head. So it must compress the output tensor of model's Transformer Encoder component to a vector of features for every data point in the current batch, in addition to a stack of Dense layers. This will be done using a pooling layer such as the GlobalAveragePooling1D layer. Then, at the Output Layer, the authors change the “Action Probability” to the “Quantile Probability”. The authors label the output with the quantiles the model is predicting.
$$\\\text{GlobalAveragePooling1D}(X) = \frac{1}{n} \sum_{i=1}^{n} X_i \quad(10)$$
$$Y_{\text{quantile}} = \text{GlobalAveragePooling1D}(X) \quad(11)$$
\paragraph{ Model setting}\mbox{}\\

\textbf{\textit{Input Layer}}: takes sequences of length 5, each with a single feature per time step, as input.

\textbf{\textit{Normalization}}: A layer normalization is applied to the input sequences to ensure stable training.

\textbf{\textit{Multi-Head Attention Blocks}}: employs four multi-head attention blocks, each followed by dropout regularization and residual connections.

\textbf{\textit{Convolutional Layers}}: following each attention block, a one-dimensional convolutional layer is applied to capture local patterns.

\textbf{\textit{Residual Connections}}: residual connections are added after each convolutional layer to facilitate information flow.

\textbf{\textit{Global Average Pooling}}: The global average pooling layer is employed to reduce spatial dimensions and capture global information.

\textbf{\textit{Dense Layers}}:Two fully connected dense layers with ReLU activation are utilized for feature extraction.

\textbf{\textit{Output Layer}}: the authors change the loss function from the MSE into the Quantile Loss.

\subsection{Experimental Design}
\subsubsection{Model Performance Evaluation }
\paragraph{Data Preparation}
In the context of data preprocessing for machine learning tasks, it is common to apply normalization techniques to the features to ensure that they contribute equally to the analysis. This paper employs min-max normalization to rescale the data features into a smaller, uniform range. The formula for min-max normalization is as follows:
$$X_{\text{normalized}} = \frac{X - X_{\text{min}}}{X_{\text{max}} - X_{\text{min}}} \quad(12)$$

where:
\textbf{\textit{$X$}} represents the original value of the feature.
\textbf{\textit{$X_min$}}is the smallest value that the feature takes on in the dataset
\textbf{\textit{$X_max$}}is the largest value of the feature in the dataset. 
\paragraph{Experiment  Construction}
The authors construct our experiments by first defining the criteria for model comparison. The models included in our testing phase are:
\begin{itemize}
\item Quantile Gradient Boosting Regression
\item Quantile Linear Regression
\item Quantile LSTM
\item Quantile TCN (Temporal Convolutional Network)
Item Quantile-Attention LSTM
\item FutureQuant Transformer
\end{itemize}
By utilizing the quantile loss function, each model is fine-tuned to predict specific points in the distribution of future market prices, providing a comprehensive understanding of potential outcomes.
\section{Experimental Results and Analysis}
The results of our experiments are visualized in the form of prediction interval plots for two of the models: the Quantile-Attention LSTM and the FutureQuant Transformer. These visualizations capture the predicted mean along with the associated prediction intervals at different quantiles for a section of the futures market data over a specific time frame.
\subsection{Quantile-Attention LSTM with Predicting Interval}\mbox{}\\
Fig. 2. illustrates the performance of the Quantile-Attention LSTM model. The shaded areas represent the prediction intervals at two quantile ranges, 0.05-0.95 and 0.1-0.9, providing insight into the model's confidence in its predictions. The intervals are overlaid on the actual last price movements, showcasing the model's ability to encapsulate potential price variations.
\clearpage
\begin{figure}[htp]
    \centering
    \includegraphics[width=12cm,height=6cm]{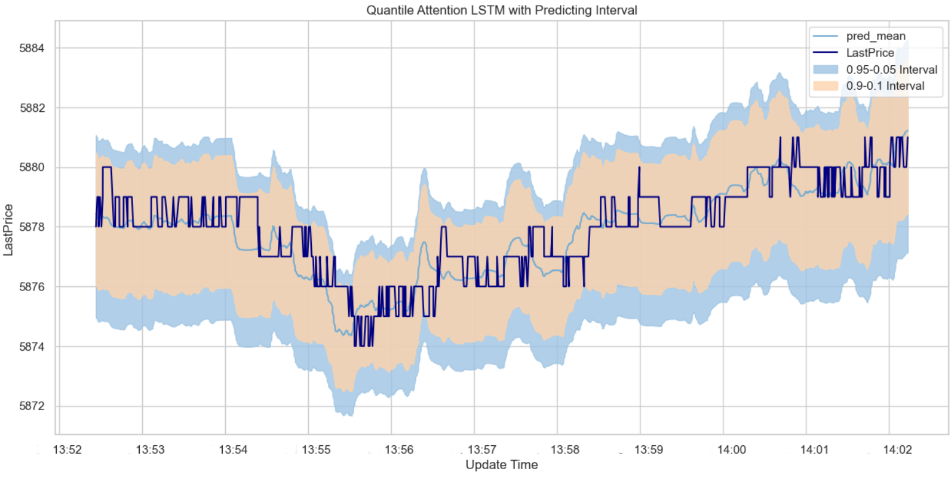}
    \caption{Quantile Attention LSTM with Predicting Interval}
    \label{fig:galaxy}
\end{figure}

\subsection{FutureQuant Transformer with Predicting Interval}\mbox{}\\
Fig. 3. displays the predictive results from the FutureQuant Transformer. Similar to the Quantile-Attention LSTM model, the prediction intervals and mean predictions are plotted against the actual price data. The narrower prediction intervals at certain time points indicate moments where the model has higher certainty in its predictions.
\begin{figure}[htp]
    \centering
    \includegraphics[width=12cm,height=6cm]{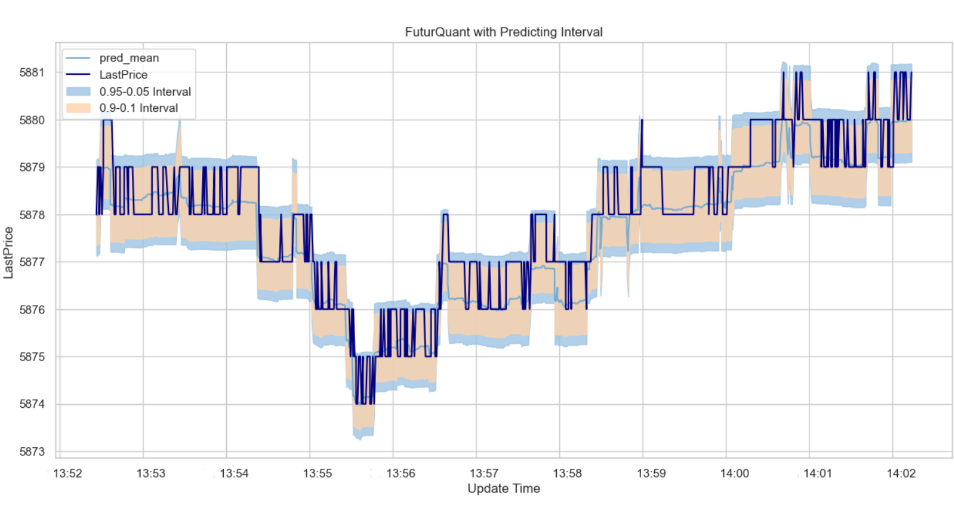}
    \caption{FutureQuant with Predicting Interval}
    \label{fig:galaxy}
\end{figure}

\subsubsection{Compared to the past observation}
\begin{figure}[htp]
    \centering
    \includegraphics[width=12cm,height=6cm]{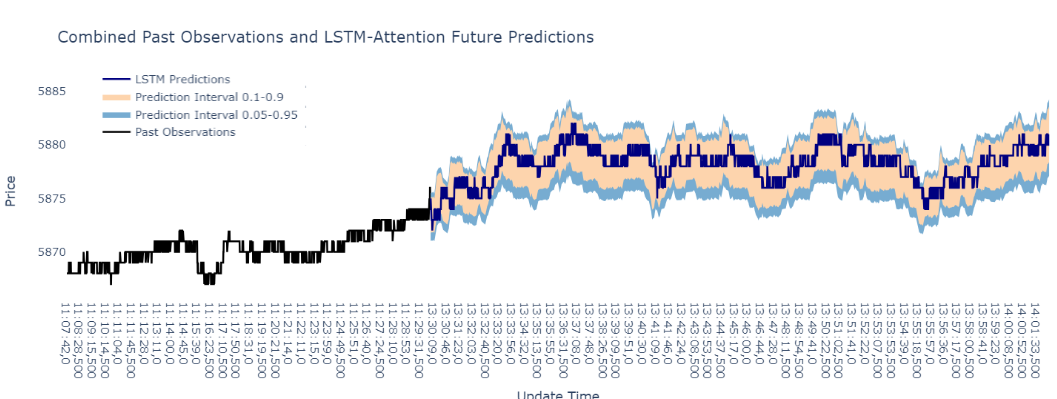}
    \hspace{2cm}
    \includegraphics[width=12cm,height=6cm]{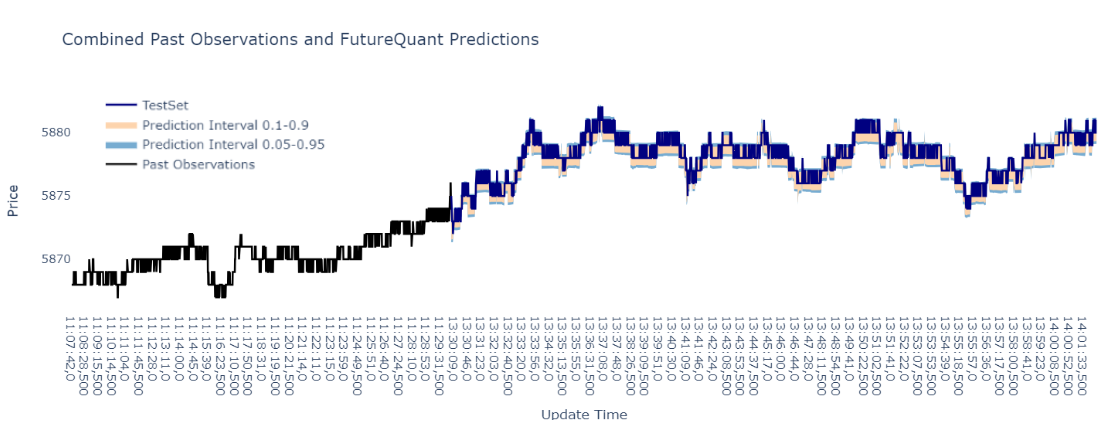}
    \caption{Combine Past Observations with Prediction Intervals (Attention-LSTM VS FutureQuant)}
    \label{fig:combined-galaxy}
\end{figure}
Fig. 4. shows the FutureQuant Transformer and the LSTM-Attention model in their ability to predict futures market prices. The predictions utilize the last available data from the morning session to forecast market trends within the day.
The FutureQuant Transformer demonstrates superior predictive performance, with the predicted price distribution closely hugging the real market values. This closeness indicates a more reliable and precise predictive capacity, which is critical for making informed trading decisions in the highly volatile futures market.
\subsection{Evaluation Metrics}
\subsubsection{Mathematical testing}
In evaluating the predictive quality of our model's Prediction Intervals (PIs), adopted two metrics to quantify their coverage and their width, recognizing that an effective PI should encapsulate the true value with a reasonably tight range. These metrics are:
\paragraph{Prediction Interval Coverage Probability (PICP)}
This measure calculates the proportion of actual values that fall within the prediction intervals over all observations. It is defined as:
$$PICP = \frac{1}{n} \sum_{i=1}^{n} c_i, \quad c_i = \begin{cases} 1, & \text{if } y_i \in [L_i, U_i] \\ 0, & \text{if } y_i \notin [L_i, U_i] \end{cases} \quad(13)$$
where \(y_i\) is the actual value, and \([L_i, U_i]\) represents the lower and upper bounds of the PI, respectively.

\paragraph{Prediction Interval Normalized Average Width (PINAW)}
This metric assesses the average width of the PIs, normalized by the range of the observed values:
$$PINAW = \frac{1}{n\Delta y} \sum_{i=1}^{n} (U_i - L_i), \quad  \Delta y
 = y_{\text{max}} - y_{\text{min}} \quad(14)$$
Here, $y_{\text{max}}$ and$y _{\text{min}}$denote the maximum and minimum observed values in the dataset, which standardize the interval width relative to the variability in the data.
While PICP indicates coverage, it is not a standalone measure of PI quality. Broader intervals tend to have high coverage, which may not be particularly informative. Thus, an optimal PI should achieve a PICP close to the desired confidence level while simultaneously minimizing PINAW.

To integrate these aspects into a singular measure of PI quality, the authors employ a modified version of the Coverage Width-based [10] Criterion (CWC), which is designed to penalize both under- and over-coverage equally:
$$CWC = (1 - PINAW) \cdot \exp\left[-\eta \cdot \left(PICP - (1-\beta)^2\right)\right]\quad (15)$$
In the equation above, $\eta$ is a user-determined parameter that balances the contribution of PINAW and PICP. For our experiments, the authors set $\eta$ to 30 (The value of 30 could be used to scale the exponential part of the equation to a particular range that is meaningful for the application). This modified CWC allows us to summarize the performance of our PIs with a single value, capturing both their precision and reliability.To assess the performance of different models, we have established an upper bound of 0.95 and a lower bound of 0.05 for the Prediction Interval Coverage Probability (PICP), ensuring a 90\% prediction interval

\textbf{here are our testing results}:
\begin{table}[htbp]
\caption{Comparison of Models}
\centering
\begin{tabular}{|c|c|c|}
\hline
\textbf{Model} & \textbf{PICP} & \textbf{CWC} \\ \hline
quantile Gradient Boosting Regression & 0.144 & 24.923 \\ \hline
quantile linear regression & 0.541 & 14.034 \\ \hline
quantile LSTM & 0.549 & 14.085 \\ \hline
quantile TCN & 1.0 & 20.609 \\ \hline
quantile-attention-LSTM & 0.998 & 7.246 \\ \hline
\textbf{FuturQuant Transformer} & \textbf{0.962} & \textbf{3.175} \\ \hline
\end{tabular}
\label{tab:model-comparison}
\end{table}
\\
The results from our experiments with the FutureQuant model demonstrate a superior coverage rate,  as indicated by the Prediction Interval Coverage Probability (PICP) of 0.962, which surpasses the targeted coverage rate of 0.90. Additionally, the model boasts the narrowest width for prediction intervals,  as reflected by the lowest Coverage Width-based Criterion (CWC) among the compared models. These metrics highlight the  model's ability to precisely capture the range of potential future market prices within exceptionally tight intervals,  thereby suggesting high prediction accuracy and an adept understanding of market volatility. The evidence of superior  performance with an optimal PICP and minimal CWC underscores the model's robustness and its potential in offering  valuable insights for trading strategies in the futures market.
\subsubsection{Trading Strategy Testing}
To test the usefulness of our trading model in guiding our trading strategy, we have used skewness, kurtosis, Bollinger band, RSI, and other indicators to create our trading strategy.

First, we created the Bollinger band indicator based on the quantiles (0.95, 0.9, 0.5, 0.1, 0.05) of the forecasts of our two quantile models, and second, we found the predicted skewness, kurtosis, and kurtosis of the futures commodity based on the five predicted prices using the least squares method. In addition, we also constructed RSI  (Relative Strength Index) and ATR (Average True Range), which are both technical analysis indicators used by traders to assess market conditions and potential price movements.

In the evaluation of financial models and trading strategies, cumulative return and drawdown are two critical metrics that encapsulate both the reward and risk dimensions. Cumulative return is a measure of total percentage gain or loss over a specific period and is instrumental in gauging the overall profitability of an investment strategy. It is computed through the formula:
$$Cumulative Return = \left( \prod_{i=1}^{n} (1 + r_i) \right) - 1 \quad(16)$$
Where \( r_i \) represents the return for period \( i \), providing a comprehensive view of an investment's growth trajectory.

RSI (Relative Strength Index): and other common indicators to build our trading algorithm, the process is as follows:
\begin{algorithm}
\caption{Generate trading signal based on technical analysis}
\label{alg:trading_signal}
\begin{algorithmic}[1]
\Function{GenerateSignal}{$price$, $ATR$, $lowerBand$, $RSI$}
    \If{$RSI < 30$} \Comment{Over-sold condition}
        \If{$price < threshold \times lowerBand$}
            \If{$ATR \geq 1\%$ \textbf{and} $ATR < 3\%$}
                \State \Return{``Buy Signal''}
            \ElsIf{$ATR \geq 3\%$}
                \State \Return{``None''}
            \EndIf
        \EndIf
    \ElsIf{$RSI > 70$} \Comment{Over-bought condition}
        \If{$price > threshold \times lowerBand$}
            \If{$ATR \geq 1\%$ \textbf{and} $ATR < 3\%$}
                \State \Return{``Sell Signal''}
            \ElsIf{$ATR \geq 3\%$}
                \State \Return{``None''}
            \EndIf
        \EndIf
    \EndIf
    \State \Return{``None''} \Comment{Default return}
\EndFunction
\end{algorithmic}
\end{algorithm}

Conversely, drawdown is a risk metric that examines the potential downside of an investment strategy. It measures the peak-to-trough decline during a specific record period of an investment portfolio, offering insight into the potential losses that could occur over the strategy's execution. A smaller drawdown is indicative of a strategy with lower risk and higher resilience to market volatility.

The superiority of a model or trading strategy is often demonstrated by its ability to achieve high cumulative returns while maintaining a low drawdown. A model that consistently yields high cumulative returns suggests a robust predictive ability and an effective strategy for realizing gains from market movements. Meanwhile, minimal drawdowns signify that the strategy effectively mitigates risks, safeguarding investments against significant declines. The combination of these two metrics provides a dual perspective, revealing not only the strategy’s capability to capitalize on profitable opportunities but also its strength in risk management.

Utilizing cumulative return and drawdown, we can quantitatively compare different models and strategies, distinguishing those that deliver sustained performance and manage risk adeptly. A strategy with a high cumulative return and a low drawdown profile is considered superior, reflecting an optimal balance between maximizing returns and controlling potential losses, which is the hallmark of effective financial management.

\begin{figure}[htp]
    \centering
    \includegraphics[width=10cm]{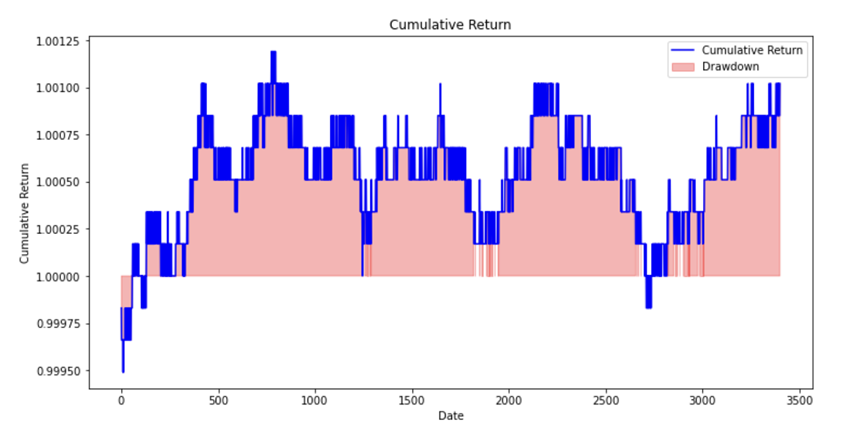}
    \caption{Cumulative return of Quantile-Attention LSTM}
    \label{fig:galaxy}
\end{figure}
\clearpage
\begin{figure}[htp]
    \centering
    \includegraphics[width=10cm]{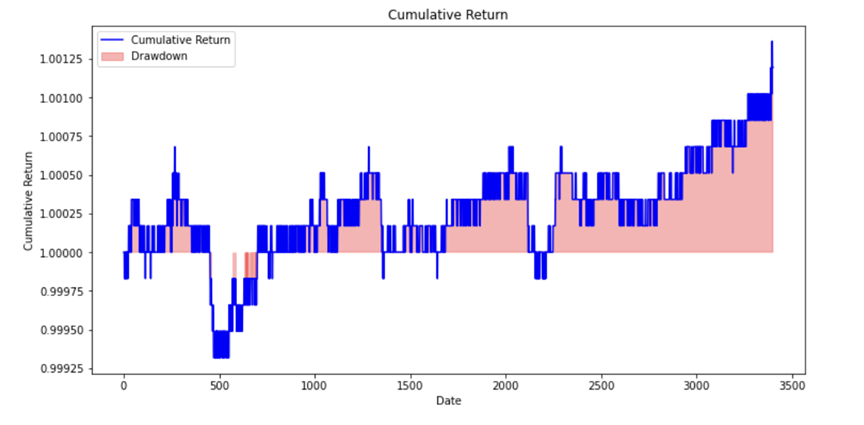}
    \caption{Cumulative return of FutureQuant}
    \label{fig:galaxy}
\end{figure}
The final cumulative return of Fig. 6. is higher than that of Fig. 5., and the maximum Drawdown is smaller than that of quantile-Attention LSTM, which indicates that the trading of Fig. 1. is more stable and less volatile, and it is not prone to black swan events, and the overall level of Drawdown is within a controllable range. The peak is reached at the last moment, while the quantile-Attention LSTM peaked in the early moments of the start of trading, and has since fluctuated and never reached the peak again. This indicates that the forecast value of FutureQuant Transformer is more accurate, so that we can judge the time to buy or sell more correctly when making trading strategies.

\begin{table}[H]
\centering
\caption{Cumulative return Comparison}
\label{tab:my_label}
\resizebox{9cm}{!}{
\begin{tabular}{|l|c|c|}
\hline
\textbf{Model}      & \textbf{Number of Drawdown \(> \)0.001} & \textbf{Cumulative Return (30mins)} \\ \hline
Quantile-Attention LSTM                & 320                                 & 0.102117\%                          \\ \hline
FutureQuant Transformer         & 193                                 & 0.1193\%                            \\ \hline
\end{tabular}
}
\end{table}

As we can see in the plot, it is obvious that the number drawdown of quantile-AttentionLSTM that is larger than 0.001 is much more than FutureQuant Transformer, which directly indicates the risk of quantile-Attention LSTM is much higher than FutureQuant Transformer, and the trading signals generated by the LSTM may not be appropriate enough. The cumulative return earned by the transformer is also higher than quantile-Attention LSTM.
\subsection{BackTesting}
The Table 3 above shows the backtesting results of the original true value data, applying the trading strategy we created to the true value to get it's true value, the blue line represents the cumulative return, and the red area represents the drawdown area, here we use FutureQuant Transformer as an example.
\begin{figure}[htp]
    \centering
    \includegraphics[width=12cm]{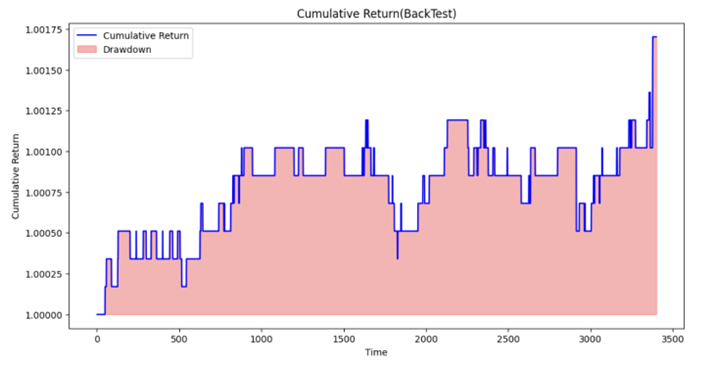}
    \caption{Backtesting result for FutureQuant Transformer}
    \label{fig:galaxy}
\end{figure}
\begin{table}[H]
\centering
\caption{Investment Performance Metrics}
\begin{tabular}{lr}
\hline
\textbf{Metric}               & \textbf{Value} \\ \hline
Cumulative Return (30mins)    & 0.175\%        \\
Cumulative Return (1-month)   & 21\%           \\
Cumulative Return (1-year)    & 264.6\%        \\
Volatility                    & 1.653    \\
Maximum Drawdown              & -35.997\%      \\ \hline
\end{tabular}
\label{tab:BackTesting result}

\end{table}
\begin{figure}[htp]
    \centering
    \includegraphics[width=12cm]{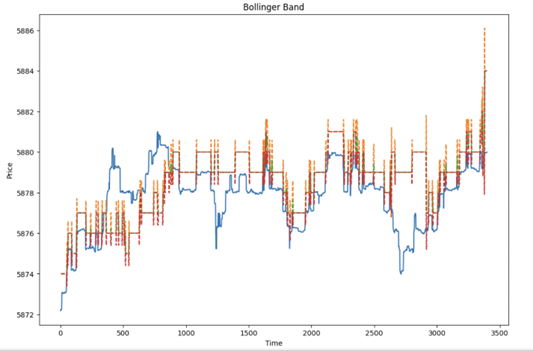}
    \caption{Backtesting result for FutureQuant Transformer}
    \label{fig:galaxy}
\end{figure}
The backtest results in this article may be a good performance, but just an example does not mean that the trading strategy for all futures is effective!
 
The essence of this trading strategy is to utilize a threshold and a series of indicators such as Bollinger Band, RSI, and ATR. Although the strategy can deliver some returns, it still has some limitations and risks due to the complexity and uncertainty of the market. For example, when the market is in a long-term trend or a major event occurs, the effectiveness of this strategy may decrease or fail, resulting in poor returns. In addition, the parameters and thresholds of this strategy need to be adjusted according to market conditions and personal experience; otherwise, it may lead to poor trading results.

\begin{table}[H]
\centering
\caption{Comparative Performance Metrics}
\resizebox{9cm}{!}{
\begin{tabular}{|l|c|c|}
\hline
\textbf{Metric} & \textbf{Quantile-Attention LSTM} & \textbf{FutureQuant Transformer} \\ \hline
Cumulative Return (30mins) & 0.102\% & 0.119\% \\ \hline
Cumulative Return (1-month) & 12.254\% & 14.316\% \\ \hline
Cumulative Return (1-year) & 154.400\% & 180.381\% \\ \hline
\end{tabular}
}

\label{table:performance}
\end{table}

\subsection{Scenario Test}  
Initial funds in account: 1,000,000\\
Return after one month: 143,160\\
Account balance after one month: 1,143,160

\begin{table}[H]
\centering
\caption{Account Balance Scenario Test}
\resizebox{9cm}{!}{
\begin{tabular}{|l|c|c|}
\hline
& \textbf{Quantile-Attention LSTM} & \textbf{FutureQuant Transformer} \\ \hline
Initial funds in account: & \$1,000,000 & \$1,000,000 \\ \hline
Account balance after one month: & \$1,122,540 & \$1,143,160 \\ \hline
\end{tabular}
}
\label{table:scenario_test}
\end{table}
Overall, this trading strategy is a simple, easy-to-understand, risk-controlled and widely applicable trading strategy with some practical application value. However, in actual trading, due to the uncertainty and complexity of the market, the effectiveness of this strategy is affected by a variety of factors and requires careful assessment and risk control. In addition, due to the changes and continuous evolution of the market, the trading strategy needs to be continuously optimized and improved to adapt to the changes in the market.

\section{Conclusion and Future Work}

\subsection{Conclusion}
The FutureQuant Transformer model exhibits outstanding performance with a Prediction Interval Coverage Probability (PICP) of 0.962 and a Coverage Width-based Criterion (CWC) of 3.175. These results underscore the model's proficiency in capturing the full spectrum of market behavior, transcending the capabilities of conventional LSTM-based models. By leveraging quantile regression, the model provides not only precise predictions but also a nuanced understanding of market uncertainties and price distributions. The integration of a quantile-attention mechanism enhances the model's precision, ensuring robustness in its forecasting ability. With its innovative approach and exceptional performance metrics, the FutureQuant Transformer establishes a new benchmark for predictive analytics, offering a sophisticated tool for navigating the volatile realm of futures trading. For the trading strategy, we use Skewness and kurtosis to recognize the market environment and risk, and then apply financial indexes to generate trading signals in a less risky way. By using quantiles of price, we leverage the accuracy of the trading strategy. The accuracy of predictions from the transformer model enables us to generate appropriate signals for trading, taking into account indicators such as the Bollinger Band, Relative Strength Index, Average True Range, and a predefined threshold level.
\subsection{Future Work}
Moving forward, enhancing the FutureQuant Transformer involves several research avenues. First, refining the network architecture to better generalize across varying market conditions may improve prediction accuracy. Additionally, investigating multi-futures quantile forecasting for diversified portfolio construction may reduce risks associated with single-commodity investments, leading to more resilient asset allocation and optimized returns. Finally, integrating the model's predictions into algorithmic trading systems offers a pathway to more efficiently capitalize on market opportunities and adapt to market changes swiftly.
\section{Author Contribution}
Wenhao Guo, Yuda Wang, and Zeqiao Huang designed the study. Wenhao Guo did the data collection and C++ API writing. Wenhao Guo and Yuda Wang did the machine learning model building. Zeqiao Huang did the trading strategy design. Wenhao Guo, Yuda Wang, and Zeqiao Huang did the backtracking and scenario test. Changjiang Zhang and Shuming Ma supervised the study. All authors reviewed and approved the final manuscript.
\newpage
\section*{References}
{
\small

[1] Zhang, Z., Zohren, S., and Roberts, S. (2018). DeepLOB: Deep Convolutional Neural Networks for Limit Order Books. *IEEE Transactions on Signal Processing*, 67(11), 3001-3012. DOI:10.1109/TSP.2019.2907260.

[2] Roa-Vicens, J., Zohren, S., and Roberts, S. (2022). Graph and Tensor-Train Recurrent Neural Networks for High-Dimensional Models of Limit Order Books. *ACM Transactions on Modeling and Computer Simulation*, 32(3), Article 23. DOI:10.1145/3533271.3561710. 

[3] Wallbridge, J. (2020). Transformers for Limit Order Books. arXiv: Computational Finance,arXiv: Computational Finance.

[4] Fischer, T., and Krauss, C. (2018). Deep learning with long short-term memory networks for financial market predictions. *European Journal of Operational Research*, 270(2), 654-669. DOI:10.1016/j.ejor.2018.02.008.

[5] R. Foygel Barber, E. J. Candes, A. Ramdas, and R. J. Tibshirani, "The limits of distribution-free conditional predictive inference," Information and Inference: A Journal of the IMA, vol. 10, no. 2, pp. 455–482, 2021

[6] Y. Romano, E. Patterson, and E. Candes, "Conformalized quantile regression," Advances in neural information processing systems, vol. 32, 2019. 

[7]  R. Koenker and G. Bassett Jr, "Regression quantiles," Econometrica: journal of the Econometric Society, pp. 33–50, 1978.

[8] J. W. Taylor, "A quantile regression neural network approach to estimating the conditional density of multiperiod returns," Journal of Forecasting, vol. 19, no. 4, pp. 299–311, 2000.

[9] S. Smyl, "A hybrid method of exponential smoothing and recurrent neural networks for time series forecasting," International Journal of Forecasting, vol. 36, no. 1, pp. 75–85, 2020.

[10] Shen, Y., Wang, X., and Chen, J. (2018). Wind Power Forecasting Using Multi-Objective Evolutionary Algorithms for Wavelet Neural Network-Optimized Prediction Intervals. Applied Sciences, 8(2), 185. https://doi.org/10.3390/app8020185

[11] Vaswani, A., Shazeer, N., Parmar, N., Uszkoreit, J., Jones, L., Gomez, AidanN., … Polosukhin, I. (2017). Attention is All you Need. Neural Information Processing Systems,Neural Information Processing Systems.

[12]Singleton, J. C., and Wingender, J. (1986). Skewness Persistence in Common Stock Returns. The Journal of Financial and Quantitative Analysis, 21(3), 335. https://doi.org/10.2307/2331046 

[13] Mei, D., Liu, J., Ma, F., and Chen, W. (2017). Forecasting stock market volatility: Do realized skewness and kurtosis help? Physica A: Statistical Mechanics and Its Applications, 481, 153–159. https://doi.org/10.1016/j.physa.2017.04.020 

[14] Lutey, M. (2023). Robust Testing for Bollinger Band, Moving Average and Relative Strength Index. Journal of Finance Issues, 20(1), 27–46. https://doi.org/10.58886/jfi.v20i1.3218

[15] Chen, S., Zhang, B., Zhou, G., and Qin, Q. (2018). Bollinger Bands Trading Strategy Based on Wavelet Analysis. Applied Economics and Finance, 5(3), 49. https://doi.org/10.11114/aef.v5i3.3079 

[16] Gonzalo, J., and Pitarakis, J. (2006). Threshold Effects in Cointegrating Relationships. Oxford Bulletin of Economics and Statistics, 68(s1), 813–833. https://doi.org/10.1111/j.1468-0084.2006.00458.x 
[17] G. Horváth. (2019). A New Indicator: Average True Range in the Context of Technical Analysis. Finance a úvěr – Czech Journal of Economics and Finance

[18] Kawakami, T. (n.d.). Quantile prediction for Bitcoin returns using financial assets’ realized measures.

[19] Jensen, V., Bianchi, F., and Anfinsen, S. (n.d.). Ensemble Conformalized Quantile Regression for Probabilistic Time Series Forecasting.

[20] Gouttes, A., Rasul, K., Koren, M., Stephan, J., and Naghibi, T. (2021). Probabilistic Time Series Forecasting with Implicit Quantile Networks. Cornell University - arXiv,Cornell University - arXiv.

[21] Gasthaus, J., Benidis, K., Wang, Y., Rangapuram, S., Salinas, D., Flunkert, V., and Januschowski, T. (2019). Probabilistic Forecasting with Spline Quantile Function RNNs. International Conference on Artificial Intelligence and Statistics,International Conference on Artificial Intelligence and Statistics.

[22] Wen, R., Torkkola, K., Narayanaswamy, B., and Madeka, D. (2017). A Multi-Horizon Quantile Recurrent Forecaster. Cornell University - arXiv,Cornell University - arXiv.

[23] Akita, R., Yoshihara, A., Matsubara, T., and Uehara, K. (2016). Deep learning for stock prediction using numerical and textual information. 2016 IEEE/ACIS 15th International Conference on Computer and Information Science (ICIS). Presented at the 2016 IEEE/ACIS 15th International Conference on Computer and Information Science (ICIS), Okayama, Japan. https://doi.org/10.1109/icis.2016.7550882

[24] Salinas, D., Flunkert, V., Gasthaus, J., and Januschowski, T. (2020). DeepAR: Probabilistic Forecasting with Autoregressive Recurrent Networks. International Journal of Forecasting, 1181–1191. https://doi.org/10.1016/j.ijforecast.2019.07.001

[25] Barez, F., Bilokon, P., Gervais, A., and Lisitsyn, N. (n.d.). Exploring the Advantages of Transformers for High-Frequency Trading A P REPRINT.

[26] Briola, A., Turiel, J., and Aste, T. (2020). Deep Learning modeling of Limit Order Book: a comparative perspective. RePEc: Research Papers in Economics - RePEc,RePEc: Research Papers in Economics - RePEc.
[27] Zhang, Z., Qin, H., Yao, L., Lu, J., and Cheng, L. (2019). Interval prediction method based on Long-Short Term Memory networks for system integrated of hydro, wind and solar power. Energy Procedia, 158, 6176–6182. https://doi.org/10.1016/j.egypro.2019.01.491

[28] Barez, F., Bilokon, P., Gervais, A., and Lisitsyn, N. (2023). Exploring the Advantages of Transformers for High-Frequency Trading.

\end{document}